# Leveraging object detection for the identification of lung cancer

**Karthick Prasad Gunasekaran[1]**

College of Information and Computer Sciences, University of Massachusetts, Amherst, United States[1]

**Abstract**: Lung cancer poses a significant global public health challenge, emphasizing the importance of early detection for improved patient outcomes. Recent advancements in deep learning algorithms have shown promising results in medical image analysis. This study aims to explore the application of object detection particularly YOLOv5, an advanced object identification system, in medical imaging for lung cancer identification. To train and evaluate the algorithm, a dataset comprising chest X-rays and corresponding annotations was obtained from Kaggle. The YOLOv5 model was employed to train an algorithm capable of detecting cancerous lung lesions. The training process involved optimizing hyperparameters and utilizing augmentation techniques to enhance the model's performance. The trained YOLOv5 model exhibited exceptional proficiency in identifying lung cancer lesions, displaying high accuracy and recall rates. It successfully pinpointed malignant areas in chest radiographs, as validated by a separate test set where it outperformed previous techniques. Additionally, the YOLOv5 model demonstrated computational efficiency, enabling real-time detection and making it suitable for integration into clinical procedures. This proposed approach holds promise in assisting radiologists in the early discovery and diagnosis of lung cancer, ultimately leading to prompt treatment and improved patient outcomes.

## I. INTRODUCTION

Because of many scientific variables, the continuous growth of technology has generated changes in people's lifestyles. Changes in lifestyle cause changes in the structure and function of human genetic cells (Deoxyribonucleic acid (DNA)) [1]. The structure-changed DNA has been partitioned into two new cells, resulting in the formation of duplicate DNA, which is utilised to replace the old DNA and dying DNA, a process known as mutation [2]. Because the mutation mechanism is influenced by numerous elements such as radiation exposure, smoking, asbestos fibre inhaling, and drinking behaviour, the erroneous mutation of a DNA cell usually results in the formation of cancer [3]. Not only males (14%), but also 13% of women throughout the United States, were affected by lung tumours [4]. Furthermore, 154,050 of the 234,030 fatalities were identified globally. According to the NNACCR study, lung cancer is one of the most dangerous illnesses, with various symptoms [5] including shortness of breath, coughing, chest discomfort, voice change, sputum colour change, and blood coughing. Furthermore, lung cancer is distinguished by self-use tiredness, joint problems, weight loss, fractures of the bones, memory loss, cachexia, which is headache, bleeding, neurological difficulty, blood clots, and face edoema [6]. Clinical doctors examined these symptoms [7] using a spirometer, which measures the amount of air in the lungs and aids in determining the presence of lung cancer through the use of several screening methods [8] such as reflex evaluation, bronch scopy, a biopsy, genetic testing, and liquid biopsy.

A blood test is also performed for predicting chest infection. The National Institute for Excellence in Healthcare gives recommendations and symptoms [9] for early lung cancer detection. Among the numerous screening methods, biopsies and bronch scopy study lung cells to predict cancer-related cells. Although a biopsy can predict lung cancer, maintaining precision as well as accuracy is challenging. As a result, a computerised tomography (CT) [10] scans is performed by transmitting X-rays through the body to examine the alterations that occur. During this procedure, a dye or liquid is administered to the chest, and pictures are captured by screening the body for 10-30 minutes. When contrasted to other imaging modalities such as MRI and PET, medical imaging demonstrates CT as an effective method during analysis owing to the sequential inspection of soft organs, tissues in the lung, and delivering relevant information about the damaged region [11]. Although screening methods has aided in the prediction of lung cancer, early diagnosis and accuracy in cancer detection are challenging to maintain. As a result, CAD [12] must be used to the clinical centre in order to produce an efficient cancer forecasting [13] system utilising an optimised and intelligent approach. The optimised lung image processing methods are used to examine the inner intricacies of the body, recover the details, extract vital information, and construct a knowledgeable system for lung cancer detection. Processing procedures [14] comprise various phases such as lung pre-processing of images, affected portion segmentation, feature extraction, and prediction of lung cancer. Among the several stages, segmentation plays an important function since it analyses every pixel in the lung picture and separates the afflicted region's linked cells, which aids in the determination of cancer and noncancerous. For region segmentation, several approaches such as fuzzy c-mean clustering, K-means clustering, Hopfield neural networks, self-organizing map, agglomatric clustering, distributed clustering, and sobel are utilised [15]. Canny edge detection algorithms are used to forecast the impacted zone. In addition, numerous optimisation approaches, such as particle swarm optimisation, genetic algorithms, ant colonies, and firefly's algorithm, are utilised to optimise the clustering process. From the segmented lung area, several characteristics [16] such as local binary patterns, and spectral, analytical, and robust features are retrieved. The Hough transform is applied to the noise portion of the continuous transform [17,18]. Furthermore, typical algorithms fail to analyse any low-quality lung picture, resulting in the extraction of erroneous features and an increase in the misclassification error rate. Taking these issues into account, this paper employs the improved profuse clustering technique (IPCT) for categorising the affected region, with the weighted mean enhancement technique effectively removing the noise image and the IPCT method obtaining the affected region without avoiding any original or normal pixels. Various spectral characteristics are extracted from the afflicted region and tested for lung cancer prediction using a deep learning instantly accomplished neural network (DITNN).

## II. RELATED WORK

This section discusses lung cancer identification technique, processes, concepts, thoughts, and processing phases of lung imaging from the perspectives of many writers. [19] discusses the analysis of the lung tumour-related affected region using positron emission tomography (PET) with computerised tomography (CT) images using a fuzzy markov random field segmentation technique. The approach computes a Gaussian distribution and probabilistic distribution function to assess the distribution of pixels in the picture. Following the computation of the pixel distribution, several characteristics are investigated in order to calculate the pixel similarity. Similar pixels create a cluster, and the impacted region is recognised inside the cluster. The system's efficiency is then evaluated utilising PET and CT images. The author developed a mechanism that assures a high dice resemblance coefficient value, such as 0.85, indicating that the fuzzy randomised markov model analyses the afflicted lung tumour region satisfactorily. [20] developed a computer-aided diagnostic (CAD) method for detecting lung cancer using CT and PET images. The author investigates several issues and approaches, such as picture segmentation and nodule identification, for optimising the lung cancer diagnosis process. During the analysis, the acquired pictures are divided into two parts, namely, training and testing images, in order to evaluate the effectiveness of the CAD system implemented. In addition, the author mentioned the shortcomings of the conventional cancer detection system since the proposed CAD method helps to fix such prediction concerns in an efficient manner. [21] discusses the creation of a lung cancer diagnosis system using a convolution neural network, which solves the problems associated with manual cancer prediction. During this process, images from CT scans are gathered and analysed using a layer of neural network that performs automatic extraction of image features, which are then processed using a deep learning process for prediction of cancer-related features using a large volume of images. The authors developed a technique that aids in decision making when analysing the patient's CT scan data. [22] used a convolution neural network approach to predict lung nodules from CT scan data. During this procedure, photos from the LIDC IDRI database are gathered and fed into the stack encoder (SAE), convolutional neural network (CNN), and deep neural network (DNN) for successful categorization of lung cancer-related features as benign or malignant. The author has developed a technique that guarantees up to 84.32% accuracy. According to the many authors' talks, CT images are reviewed using segmentation and an optimised machine learning approach for accurate cancer prediction. Chronic illness diagnosis is challenging owing to an absence of data and the complexity of diagnosing. [23] study presented a deep learning-based system for computer-aided detection using three models (CAD3), one for detection, one for classification, and one for visualisation. The system is based on the YOLO v2 algorithm, which is based on the Convolutional Neural Network (CNN). The YOLO v2 approach is one of the simplest, adding additional layers to a pre-trained CNN to recognise the disease in images. [24] proposed a Yolov2 method which have Images of 2450 normal thyroid nodules and 2557 cancerous thyroid nodules were gathered and labelled, and a deep learning system for automated recognition of images and diagnosis was built utilising the YOLOv2 neural network model. The system's effectiveness in the identification of nodules in the thyroid was tested, and the utility of machine learning in clinical practise was studied. The early diagnosis of nodules in the lungs is extremely important in the prevention of lung cancer. Deep learning's target identification approach is currently frequently employed for healthcare image processing. The research in [25] offers a lung nodule detection technique based on enhanced yolov3 to increase the effectiveness and precision of lung nodule detection, as well as to minimise the missed diagnosis and misdiagnosed rate of pulmonary nodule. The experimental findings demonstrate that the modified model's average accuracy has increased between 70.5% to 73.9%, and its convergence effect is superior to that of the baseline model. The objective of [26] study is to develop a lung nodule detector that has a reasonable balance of efficiency and effectiveness and can be used immediately in the hospital setting. Authors break assignment in two steps. First, by integrating several current techniques—the depthwise over-parameterized convolutional layer, a convolutional block attention module, and the focal loss function—we primarily try to increase the model's accuracy. Finally, used redundant channel pruning to the created model to create the YOLOv4 pulmonary nodule detector, which is more effective. One of the main cancer forms with a high death rate that has paralyzed the world is lung cancer. Early detection greatly improves the prognosis for survival. The lung (pulmonary) nodules from the lung CT scan are detected with great sensitivity and precision using a customised deep learning YOLOv3[27].

## III. METHODOLOGY

A. Dataset Acquisition
Dataset acquired which is lung cancer dataset from Kaggle comprising 1500 x-ray images. This dataset includes annotations tailored for training a YOLOv5 model. Split the dataset into three subsets: 300 images were allocated for validation, 1000 images were assigned for training, and the remaining images were utilized for testing the model's performance. Some of the sample images of the dataset can be seen in the following Fig 1.

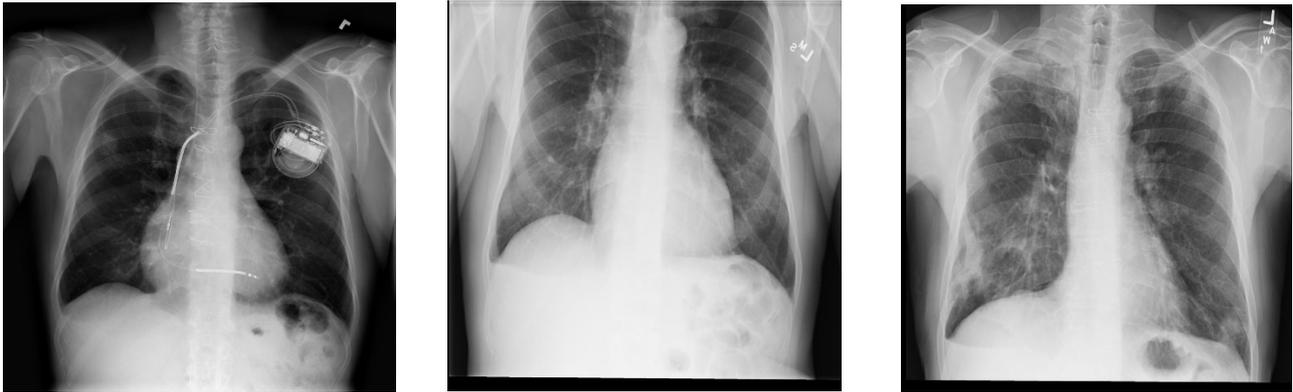

Fig. 1 Samples of Dataset

B.   Pre-processing

The x-ray images had been pre-processed in order to train the YOLOv5 model. To achieve consistency, the pre-processing stages included scaling the images to a constant input size, often in a range of 416x416 pixels. The images were also normalised to improve the model's capacity to learn and generalise. Any essential data augmentation methods, including random cropping, a rotation, or flipping it, were used to supplement and diversify the dataset. The processed x-ray images can then be utilised to train the YOLOv5 model.

C.   Model Training

To detect lung nodules in X-rays, the "Chest-ray nodule detection"-dataset was taken from Kaggle Public Domain Dedication. The total number of images from X-rays was 1500, which we then divided for training and evaluation. 1000 X-rays nodule positive images, were utilised for training, and 300 X-ray images were used for evaluation. This approach for generating synthesised data has already been discussed [18]. The detection technique is YOLOv5s, a compound-scaled detection of objects model learned on COCO datasets that combines model assembly and hyperparameter. It is divided into three major sections: Backbone module is used for extracting features from input images. CSPNet is used in the Backbone to gather features from pictures used as input images. The Neck serves to create the pyramid feature. It aids the module in determining the scaling factor of observed items of the same sort but at various scales. PANet is a method used to generate pyramid features. The module Neck serves to create pyramid features for generalisation, whereas the module Head is utilised for detection.

The main role of the head unit is to apply anchors of varying sizes to the features created in the preceding layers, as well as a bounding box with a score. SPP is an abbreviation for Spatial Pyramids Pooling.Then used a batch size of 32 to train the YOLOv5 algorithm on a lung cancer dataset. The process of training was used 50 epochs, enabling the model to learn and enhance its performance repeatedly. During the optimisation phase, a rate of learning of 0.002 was utilised to calculate the step size. This parameter influences how quickly the model responds to the data used for training. It was feasible to strike an appropriate balance between speedy convergence and avoiding becoming trapped in inferior solutions by modifying the learning rate. The whole architecture of yolov5 training can be seen in the following Fig 2.

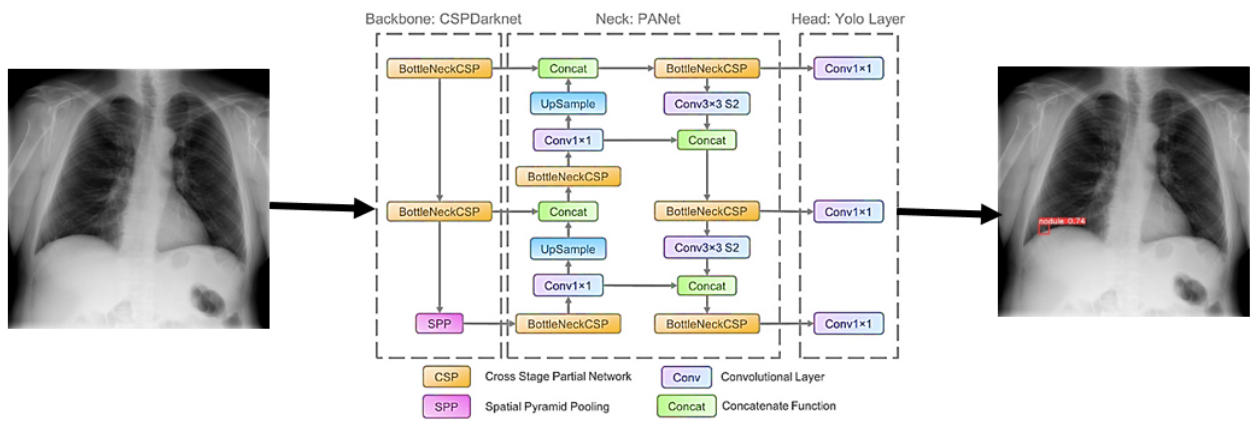

Fig. 2 Model Architecture and Training

IV.   RESULTS

A. Test Data

To evaluate the trained algorithm, the next 100 X-rays from a test dataset containing data from 100 patients were utilized. The algorithm was applied to these X-rays, and the resulting outcomes are presented below.

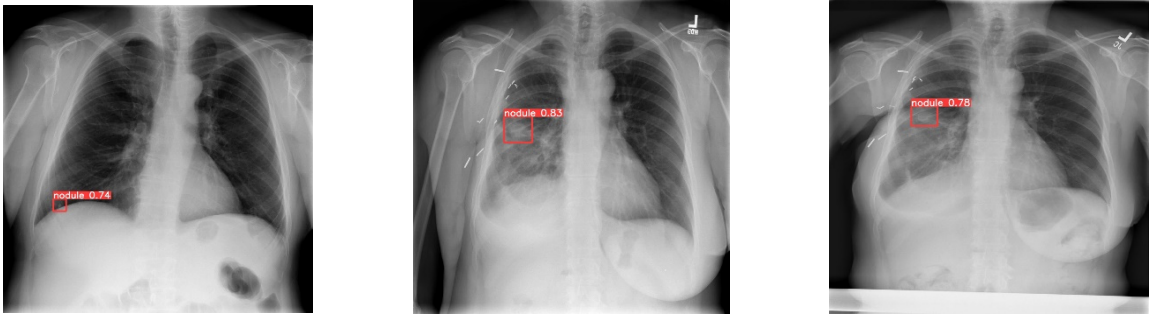

Fig. 3 Testing of Model on test Data

B. Evaluation Matrices

Biological specificity and sensitivity were evaluated using X-ray images taken from a dataset used in the follow-up treatment for chest X-rays confirming the existence of distant metastases. To produce a negative control, X-ray images unconnected to tumours but linked to trauma and non-malignant lung illness were gathered on a daily basis from an emergency hospital. It's worth noting that these images weren't utilised during the training of our yolov5 but were only used for external validation. The algorithm recognised all nodules having a test batch estimate greater than 0.5.

As assessment measures, F1 score, ROC (Receiver Operating Characteristic), and accuracy were used. This can be seen in Fig 4. The effectiveness of the system review produced encouraging findings. The accuracy in properly identifying positive instances, or sensitivity, was found to be 94%. This shows that the algorithm is quite good at accurately identifying the existence of positive cases. However, the specificity, which measures the capacity to accurately identify negative situations, was found to be 90.5%, underscoring an excellent precision in accurately recognising negative occurrences. It was discovered that the precision, which measures the percentage of accurately detected positive instances among all positive forecasts, was 100%, demonstrating that the system had a low rate of erroneous positive predictions. The system has a strong capacity to identify real positive instances, as evidenced by the recall, referred to be the true positive percentage, which was found to be 95%. Additionally, the system's overall accuracy was determined to be 91%, indicating a high level of accuracy in categorising scenarios that are both beneficial and detrimental. A favourable balance between accurately detected positive instances and false positives was found to exist, as measured by the precision-recall value, and which offers an aggregate measure of accuracy and recall, which was found to be 83%. Additionally, the F1 score, which highlights an equal efficiency between precision and recall, was discovered to be 77%, an estimation of accuracy for the system that considers both precision and recall. These metrics show the system's efficacy and dependability by showing good performance in properly recognising positive situations while keeping a low percentage of false positives.

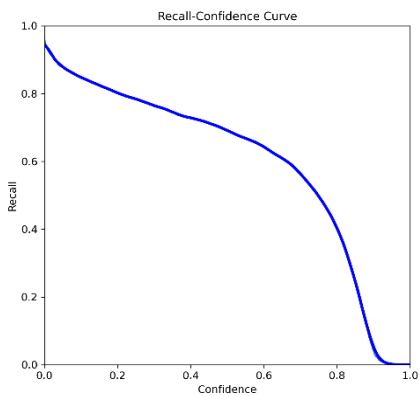
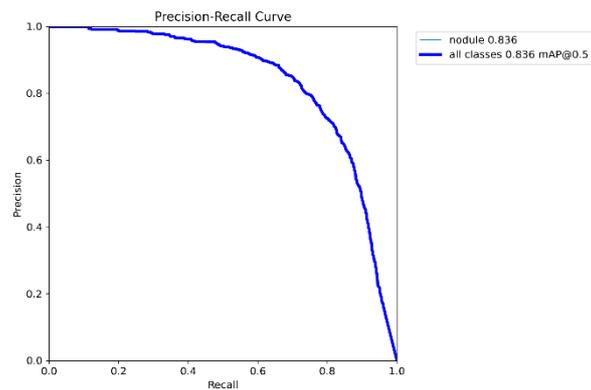

(a)            (b)

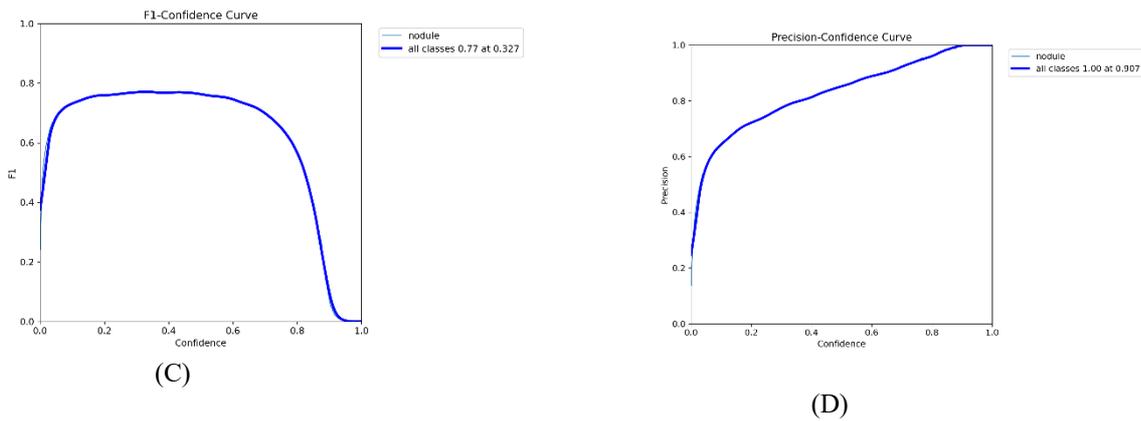

(C)                                                      (D)

Fig. 4 Evaluation Matrices

## V. CONCLUSION

There has been a tremendous increase in the field of study that investigates the application of computational intelligence in the development of diagnostics and prognostic tools in recent years. These AI-powered solutions have the ability to transform many parts of healthcare by increasing the precision, effectiveness, and availability of medical evaluations. There are currently no established applications that leverage AI technology for sarcoma lung cancer follow-up calls, which relate to the continual surveillance of individuals who were previously confirmed to have sarcoma lung cancer. This implies that traditional follow-up procedures may not fully utilise AI's potential to improve sarcoma patient assessment and care. The use of convolutional neural network models for assessing chest X-rays is one area of special attention in sarcoma follow-ups. In sarcoma follow-ups, chest X-rays are often performed to identify any symptoms of metastatic disease (the spread cancer cells to other regions of the body). The method of analysing X-rays of the chest can be enhanced by including CNN-assisted assessment. CNN algorithms. are deep learning networks that specialise in image analysis and excel in detecting patterns and anomalies in medical images.

However, while the prospective benefits of CNN-assisted assessment in cancer follow-ups are encouraging, further study is required to test and establish the systemic benefits of this hybrid method. This includes doing extensive research to assess the precision, effectiveness, and medical impact of applying CNNs in cancer follow-up methods. Furthermore, research efforts should strive to examine the combination of CNN-assisted assessment with existing diagnostic procedures, as well as its cost-effectiveness and smooth incorporation into clinical practise.Finally, the implementation of based on artificial intelligence diagnostics and prognosis tools, like CNN-assisted examination of chest X-rays, has the potential to greatly enhance results and quality of treatment for cancer patients throughout follow-up periods through thorough study and validation.

## REFERENCES

[1]. Prevalence of beliefs about actual and mythical causes of cancer and their association with socio-demographic and health-related characteristics: Findings from a cross-sectional survey in England.
[2]. Tamara Baker, Early cancer detection behaviors among black males, J. Mens Health 14 (3) (2018), ISSN: 1875-6859.
[3]. 3] Z. Liu, J. Wang, Z. Yuan, B. Zhang, L. Gong, L. Zhao, P. Wang, Preliminary results about application of intensity-modulated radiotherapy to reduce prophylactic radiation dose in limited-stage small cell lung cancer, J. Cancer 9 (15) (2018) 2625–2630, https://doi.org/10.7150/jca.24976.
[4]. Catharina Balmelli, Nikola Railic, Marco Siano, Kristin Feuerlein, Richard Cathomas, Valerie Cristina, Christiane Güthner, Stefan Zimmermann, Sabine Weidner, Miklos Pless, Frank Stenner, Sacha I. Rothschild, Lenvatinib in advanced radioiodine-refractory thyroid cancer - a retrospective analysis of the Swiss Lenvatinib named patient program, J. Cancer 9 (2) (2018) 250–255, https://doi.org/10.7150/jca.22318.
[5]. R. Manser, A. Lethaby, L.B. Irving, C. Stone, G. Byrnes, M.J. Abramson, D. Campbell, Screening for lung cancer, Cochrane Database Syst. Rev. 6 (6) (June 2013), https://doi.org/10.1002/14651858.CD001991.pub3 CD001991.
[6]. M. Usman Ali, J. Miller, L. Peirson, D. Fitzpatrick-Lewis, M. Kenny, D. Sherifali, P. Raina, Screening for lung cancer: a systematic review and meta-analysis, Preventive Med. 89 (August 2016) 301–314, https://doi.org/10.1016/j. ypmed.2016.04.015.
[7]. A. Gutierrez, R. Suh, F. Abtin, S. Genshaft, K. Brown, Lung cancer screening, Semin. Interventional Radiol. 30 (2) (June 2013) 114–120, https://doi.org/ 10.1055/s-0033-1342951. PMC 3709936.
[8]. P.B. Bach, J.N. Mirkin, T.K. Oliver, et al., Benefits and harms of CT screening for lung cancer: a systematic review, JAMA: J. Am. Med. Assoc. 307 (22) (June 2012) 2418–2429, https://doi.org/10.1001/jama.2012.5521.
[9]. D.R. Aberle, F. Abtin, K. Brown, Computed tomography screening for lung cancer: has it finally arrived? Implications of the national lung screening trial, J. Clin. Oncol. 31 (8) (March 2013) 1002–1008, https://doi.org/10.1200/ JCO.2012.43.3110.

[10]. L. Frank, L.E. Quint, Chest CT incidentalomas: thyroid lesions, enlarged mediastinal lymph nodes, and lung nodules, Cancer Imaging 12 (1) (March 2012) 41–48, https://doi.org/10.1102/1470-7330.2012.0006.

[11]. N. Murray, A.T. Turrisi, A review of first-line treatment for small-cell lung cancer, J. Thorac. Oncol. 1 (3) (March 2006) 270–278, https://doi.org/10.1016/ s1556-0864(15)31579-3.

[12]. Emre Dandıl, A computer-aided pipeline for automatic lung cancer classification on computed tomography scans, J. Healthcare Eng. 2018 (2018) 12, https://doi.org/10.1155/2018/9409267, Article ID 9409267.

[13]. G. Manogaran, P.M. Shakeel, A.S. Hassanein, M.K. Priyan, C. Gokulnath, Machine-learning approach based gamma distribution for brain abnormalities detection and data sample imbalance analysis, IEEE Access (2018 Nov 9), https://doi.org/10.1109/ACCESS.2018.2878276.

[14]. P.M. Shakeel, A. Tolba, Zafer Al-Makhadmeh Al-Makhadmeh, Mustafa Musa Jaber, Automatic detection of lung cancer from biomedical data set using discrete AdaBoost optimized ensemble learning generalized neural networks, Neural Comput. Appl. (2019) 1–14, https://doi.org/10.1007/s00521-018- 03972-2.

[15]. N. Gupta, D. Gupta, A. Khanna, P.P. Rebouças Filho, V.H.C. de Albuquerque, Evolutionary algorithms for automatic lung disease detection, Measurement 140 (2019) 590–608, https://doi.org/10.1016/j.measurement.2019.02.042.

[16]. E.O. Bursalıoğlu, F.A. Alkan, Ü.B. Barutçu, M. Demir, Y. Karabul, B. Balkan, O. Içelli, Prediction of electron density and trace element concentrations in human blood serum following radioiodine therapy in differentiated thyroid cancer patients, Measurement 100 (2017) 19–25, https://doi.org/10.1016/j. measurement.2016.12.035

[17]. S. Dutta, S.K. Pal, R. Sen, Progressive tool flank wear monitoring by applying discrete wavelet transform on turned surface images, Measurement 77 (2016) 388–401, https://doi.org/10.1016/j.measurement.2015.09.028.

[18]. J. Valença, D. Dias-da-Costa, E.N.B.S. Júlio, H. Araújo, H. Costa, Automatic crack monitoring using photogrammetry and image processing, Measurement 46 (1) (2013) 433–441, https://doi.org/10.1016/j.measurement. 2012.07.019.

[19]. H. Cui, X. Wang, D. Feng, Automated localization and segmentation of lung tumor from PET-CT thorax volumes based on image feature analysis, in: Proceedings of the 34th Annual International Conference of the IEEE Engineering in Medicine and Biology Society (EMBS '12), San Diego, Calif, USA, September 2012, pp. 5384–5387.

[20]. N. Zhang, S. Ruan, S. Lebonvallet, Q. Liao, Y. Zhu, Kernel feature selection to fuse multi-spectral MRI images for brain tumor segmentation, Comput. Vision Image Understanding 115 (2) (2011) 256–269.

[21]. Yu Guo, Yuanming Feng, Jian Sun, et al., Automatic lung tumor segmentation on PET/CT images using fuzzy markov random field model, Comput. Math. Methods Med. 2014 (2014) 6, https://doi.org/10.1155/2014/401201, Article ID 401201.

[22]. Ayman El-Baz, Garth M. Beache, Georgy Gimel'farb, et al., Computer-aided diagnosis systems for lung cancer: challenges and methodologies, Int. J. Biomed. Imaging 2013 (2013) 46, https://doi.org/10.1155/2013/942353, Article ID 942353.

[23]. Abas, Shakir Mahmood, A. Mohsin Abdulazeez, and D. Q. Zeebaree. "A YOLO and convolutional neural network for the detection and classification of leukocytes in leukemia." *Indones. J. Electr. Eng. Comput. Sci* 25.1 (2021).

[24]. Wang, Lei, et al. "Automatic thyroid nodule recognition and diagnosis in ultrasound imaging with the YOLOv2 neural network." World journal of surgical oncology 17.1 (2019): 1-9.

[25]. Joseph Redmon, Ali Farhadi, et al. " YOLOv3: An Incremental Improvement." 2018 arXiv:1804.02767, 2018.

[26]. Alexey Bochkovskiy et al. "YOLOv4: Optimal Speed and Accuracy of Object Detection." *2020 arXiv:2004.10934*, 2020.

[27]. Liu, Chenyang, et al. "Automatic detection of pulmonary nodules on CT images with YOLOv3: development and evaluation using simulated and patient data." Quantitative Imaging in Medicine and Surgery 10.10 (2020): 1917